# INTERNAL GRAVITY WAVES

# IN STRATIFIED HORIZONTALLY INHOMOGENEOUS MEDIA

Vitaly V. Bulatov (1), Yuriy V. Vladimirov (1), Vasily A. Vakorin (2)

(1) Institute for Problems in Mechanics, Russian Academy of Sciences, Moscow, Russia

(2) University of Saskatchewan, Saskatchewan, Canada.

## Abstract

The principal parameter determining the behavior of internal gravity waves is the Brunt-Vaisala frequency $N^2 = -g\,\partial \ln \rho / \partial z$ ( $g$ is the free-fall acceleration, and $\rho$ is the undisturbed density). In reality, the density $\rho$ depends not only on the vertical variable $z$ but also on the horizontal variables $x, y$. Therefore it is important to consider the problem of the internal waves excited by a point source, for example, a mass source moving in a medium with density $\rho$ (and hence $N^2$) varying with respect to all the variables. The difficulty of this problem consists in the fact that when $\rho = \rho(z, x, y)$ the partial differential equations describing the internal waves do not admit the separation of the variables. However, the fact that the characteristic horizontal scale of variation of the density is large as compared with the characteristic lengths of the internal waves makes it possible to solve the problem by means of an approximate method analogous to the method of geometric optics (ray method). The far field of internal wave is a sum of modes each of which is enclosed within its own Mach cone, the asymptotic form of each mode near the corresponding wave front being expressed in terms of certain special functions (the Airy function and its derivative, Fresnel integrals), whose argument depends on the first two coefficients of the corresponding Taylor expansion of the dispersion curves at zero (local asymptotics with the use of the weak-dispersion approximation). Within the Mach cone, i.e., far from the wave front, the asymptotic form of each mode, determined by the stationary phase method, is a locally sinusoidal wave. In the same studies the global asymptotics - global Airy and Fresnel waves - were also constructed. Therefore in investigating the problem of the internal wave field excited by a source moving in a stratified horizontally inhomogeneous medium the solution must be sought not in the form of the usual WKB expansion in locally harmonic waves but in the form of waves of a special type - global Airy and Fresnel waves.. In this paper we examine the far internal wave field at points both close to and remote from the wave front.

We will consider the problem of the propagation of the global Airy and Fresnel internal waves generated by a point mass source moving with the velocity $V$ along the $x$ axis at a depth $z_0$ in a layer $-h < z < 0$ of stratified horizontally inhomogeneous fluid with density $\rho = \rho(z, \varepsilon x, \varepsilon y)$ (the small parameter e characterizes the "slow variables," slowness of variation with respect to the variable $z$ is not assumed). We will assume that in a medium with a horizontally inhomogeneous density field it is possible to neglect the steady flows caused by that field. Then from the linearized system of equations of hydrodynamics in the Boussinesq approximation we can obtain

$$\frac{\partial^2}{\partial t^2}\left(\frac{\partial^2}{\partial x^2}+\frac{\partial^2}{\partial y^2}+\frac{\partial^2}{\partial z^2}\right)w - \frac{g}{\rho}\left(\frac{\partial^2}{\partial x^2}+\frac{\partial^2}{\partial y^2}\right)(\mathbf{v}\,\mathrm{grad}\,\rho) = \delta''_{tt}(x-Vt)\,\delta(y)\,\delta'(z-z_0) \quad (1)$$

$$\left(\frac{\partial^2}{\partial x^2}+\frac{\partial^2}{\partial y^2}\right)\mathbf{u} + \nabla\frac{\partial w}{\partial z} = \delta(z-z_0)\,\nabla[\delta(x-Vt)\,\delta(y)]\,, \quad \nabla = (\partial/\partial x, \partial/\partial y)\,, \quad \mathbf{v}=(\mathbf{u},\,w) \quad (2)$$

where $w$ is the vertical component of the internal wave velocity, and $\mathbf{u}$ is the horizontal velocity vector. The boundary conditions are taken in the form: $w=0$ when $z=0,\,-h$. We will seek the solution of (1), (2) in the form of a sum of modes [1-3] and relate all the calculations to the individual mode in question, omitting the index. Starting from the structure of the global asymptotics for a horizontally homogeneous medium, we seek the solution of (1), (2) in the form

$$w = A(\varepsilon x, \varepsilon y, z, \varepsilon t)F_0(\sigma) + \varepsilon^a B(\varepsilon x, \varepsilon y, z, \varepsilon t)F_1(\sigma) + O(\varepsilon^{2a}) \quad (3)$$

$$\mathbf{u} = \mathbf{u}_0(\varepsilon x, \varepsilon y, z, \varepsilon t)F_1(\sigma)\varepsilon^{a-1} + O(\varepsilon^{2a-1})\,, \quad F'_{m+1}(\sigma) = F_m(\sigma)\,, \quad \sigma \equiv \left(\frac{S(\varepsilon x, \varepsilon y, \varepsilon t)}{a\varepsilon}\right)^a \quad (4)$$

where the argument $\sigma$ is assumed to be of the order of unity, and the functions $S$, $A$, $\mathbf{u}_0$ are subject to determination. The value of $a$ is equal to $2/3$ for an Airy wave and $a = 1/2$ for a Fresnel wave. If we seek the solution for the vertical component of the Airy wave velocity, then as $F_0(\sigma)$ we must take the derivative Airy function $\mathrm{Ai}'(\sigma)$, while for the rise of the Airy wave as $F_0(\sigma)$ we must take the Airy function $\mathrm{Ai}(\sigma)$. For the vertical component of the Fresnel wave velocity as $F_0(\sigma)$ we will take the function $\Phi'(\sigma)$ and, accordingly, for the rise of the Fresnel wave [1-3]

$$F_0(\sigma) = \Phi(\sigma), \quad \Phi(\sigma) = \mathrm{Re}\int_0^\infty \exp\left(-it\sigma - i\frac{t^2}{2}\right)dt$$

In what follows, to be specific, we will consider the vertical component of the Airy wave velocity and the rise of the Fresnel wave, since by virtue of the linearity of the problem it is possible to go over from the rise $\eta$ to the vertical velocity component $w$ by setting $w$ everywhere equal to $\partial\eta/\partial t$.

Substituting (4) in (2), we find

$$\mathbf{u}_0 = -A'_z \left(\frac{S}{a}\right)^{1-a} \frac{\nabla S}{\varepsilon\left(\left(\frac{\partial S}{\partial x}\right)^2 + \left(\frac{\partial S}{\partial y}\right)^2\right)} \tag{5}$$

We substitute (3)-(5) in Eq. (1) and, equating terms with like powers of $\varepsilon$, we obtain for $\varepsilon^a$

$$\frac{\partial^2 A}{\partial z^2} + \mathbf{k}^2\left(\frac{N^2(z,x,y)}{\omega^2} - 1\right)A = 0, \qquad A = 0, \quad z = 0, -h \tag{6}$$

where we have introduced the following notation: $\mathbf{k} \equiv (p,q) = \nabla S$, $\omega = \partial S/\partial t$. In solving the problem we assume that the dispersion relation, denoted by $K(\omega, x, y)$, is known; then for determining the function $S$ we have the eikonal equation

$$\left(\frac{\partial S}{\partial x}\right)^2 + \left(\frac{\partial S}{\partial y}\right)^2 = K^2(\omega, x, y) \tag{7}$$

Equation (7) is the Hamilton-Jacobi equation with Hamiltonian $H(\omega, \mathbf{k}, x, y) \equiv |\mathbf{k}|^2 - K^2(\omega, x, y)$. The characteristic system for the eikonal equation (7) has the form

$$\begin{aligned}
\frac{dx}{d\tau} &= \frac{p}{K(\omega, x, y) K'_\omega(\omega, x, y)}, & \frac{dy}{d\tau} &= \frac{q}{K(\omega, x, y) K'_\omega(\omega, x, y)} \\
\frac{dp}{d\tau} &= \frac{K'_x(\omega, x, y)}{K'_\omega(\omega, x, y)}, & \frac{dq}{d\tau} &= \frac{K'_y(\omega, x, y)}{K'_\omega(\omega, x, y)}, & \frac{d\omega}{d\tau} &= 0
\end{aligned} \tag{8}$$

It is convenient to assign the initial conditions for the system (8) in the three-dimensional space $x, y, t$ on a certain surface: $t = t_0$, $x = x_0(\ell)$, $y = y_0(\ell)$. Let the eikonal $S$ be known on this surface: $S(x, y, t) = S_0(\ell, t_0)$. This corresponds to assigning the initial eikonal on a certain fixed line $x = x_0(\ell)$, $y = y_0(\ell)$ at an arbitrary moment of time $t_0$. Differentiating the initial eikonal with respect to $\ell, t_0$, we obtain a system of equations for determining the initial values of the frequency $\omega_0$ and the wave vector $(p_0, q_0)$

$$\omega_0(\ell, t_0) = \frac{\partial S_0}{\partial t_0}$$

$$p_0(\ell,t_0)x_0'(\ell) + q_0(\ell,t_0)y_0'(\ell) = \frac{\partial S_0}{\partial t_0}$$

$$p_0^2(\ell,t_0) + q_0^2(\ell,t_0) = K^2(\omega_0(\ell,t_0), x_0(\ell), y_0(\ell))$$

Thus, in the three-dimensional space $x, y, t$ the solution of system (8) defines a family of spacetime rays $x = x(t, t_0, \ell)$, $y = y(t, t_0, \ell)$, where $x = x_0(\ell)$, $y = y_0(\ell)$ at $t = t_0$. In this case $\ell$ and the moment of departure of the ray $t_0$ play the part of ray coordinates, and the variable $t$ is simultaneously a Cartesian and a ray coordinate. As distinct from the characteristic system for a monochromatic wave, the projections of the spacetime rays on the plane $x, y$ define a two-parameter family of rays, which for fixed $t_0$ goes over into the usual one-parameter family of rays with ray coordinates $\ell$ and $t$. Nevertheless, in what follows we will call the variables $\ell$ and $t_0$ the ray coordinates. As may be seen from (8), along the ray the frequency $\omega$ preserves its initial value $\omega_0(\ell, t_0)$ and the eikonal $S^*(t, t_0, \ell) = S(x(t, t_0, \ell), y(t, t_0, \ell), t)$ in ray coordinates is determined by integration along the ray:

$$S^*(t, t_0, \ell) = S_0(\ell, t_0) + \omega_0(\ell, t_0)(t - t_0) + \int_{t_0}^{t} \frac{K(\omega_0(\ell, t_0), x(\tau, t_0, t), y(\tau, t_0, t))}{K_\omega'(\omega_0(\ell, t_0), x(\tau, t_0, t), y(\tau, t_0, t))} d\tau$$

In order to find $S(x, y, t)$ and $\omega(x, y, t)$ in the Cartesian coordinates $x, y, t$, it is sufficient to invert the ray equations $\ell = \ell(t, x, y)$, $t_0 = t_0(t, x, y)$. For this it is necessary that for any $t$ the Jacobian $D(t, t_0, \ell) \equiv x_{t_0}' y_\ell' - x_\ell' y_{t_0}'$ be nonzero.

In order to determine the amplitude function $A$, we first represent $A(x, y, z, t)$ in the form $A(x, y, z, t) = A^*(x, y, z, \omega(x, y, t)) = \psi(x, y, \omega) f(x, y, z, \omega)$, where $f$ is the normalized eigenfunction of the problem (6)

$$\int_{-h}^{0} [N^2(z, x, y) - \omega^2] f^2(x, y, z, \omega) dz = 1$$

Then, substituting (3)-(5) in (1), using the properties of the Airy function and Fresnel integrals, and equating the terms with $\varepsilon^{2a}$, after some cumbersome calculations for the Airy wave we obtain

$$\frac{d}{dt}\psi^2 P + \nabla^* \mathbf{c} = 0 \qquad (9)$$

$$\nabla^* = \nabla + \frac{\partial}{\partial \omega}\nabla\omega, \qquad \frac{d}{dt} = \frac{\partial}{\partial t} + \frac{\mathbf{k}\nabla}{K(\omega,x,y)K'_\omega(\omega,x,y)},$$

$$P = \sqrt{\sigma(\omega,x,y)}\, K'_\omega(\omega,x,y)\, K^{-2}(\omega,x,y)$$

where $\mathbf{c}$ is the internal wave group velocity vector, and $d/dt$ is the derivate along the (8). For the Fresnel wave $P = K'_\omega(\omega,x,y)$. We then use the Liouville theorem, which states that along the characteristics (8) the relation $dD(t,t_0,l)/dt = \nabla^* \mathbf{c}$ will hold. Then (9) can be written in the form of a certain law of conservation along the ray:

$$\frac{d}{dt}\ln(D\psi^2 P) = 0 \qquad (10)$$

From this, assuming that all the functions depend on the ray coordinates, we obtain

$$D(t,t_0,\ell)\,\psi^2(t,t_0,\ell)\,P(t,t_0,\ell) = C(\ell,t_0) \qquad (11)$$

where $C(\ell,t_0)$ is a function which cannot be determined by solving the problem by the method of geometric optics. We note that, by analogy with the case of locally harmonic waves, the conservation law (10) can be written in the form:

$$\frac{DI}{Dt} + \nabla^*(I\mathbf{c}) = 0, \qquad \frac{D}{Dt} = \frac{\partial}{\partial t} + \frac{\partial}{\partial \omega}\frac{\partial \omega}{\partial t}$$

where $I = \psi^2 P$ is the "adiabatic invariant" of the corresponding wave.

For the final solution of the problem it is necessary to determine the function $C(\ell,t_0)$ in Eq. (11), which can be found by using the solution of the problem of the motion of a point mass source in a stratified horizontally homogeneous medium [1-3], since at typical distance from the source, at which the global asymptotics in a horizontally homogeneous medium are valid (of the order of several $h$), it may be assumed that the parameters of the medium characterizing the horizontal inhomogeneity vary little, i.e., the medium may be assumed to be locally homogeneous in the horizontal direction. At the moment of time $t = t_0$ let the source, moving with the velocity $V$, be located at the point $(x_0, y_0) = (Vt_0, 0)$. At each moment of time $t_0$ the source radiates waves of all frequencies on the interval $0 < \omega < \max N(z)$; therefore as the ray coordinates it is convenient to take $\omega$ and the moment of departure of the ray from the source $t_0$. Then for the Airy wave we obtain [1-3]

$$C(\omega, t_0) = \frac{\omega^4}{2 V K^3(\omega, x_0, y_0) v(\omega, x_0, y_0)} \left( \frac{\partial f(x_0, y_0, z_0, \omega)}{\partial z_0} \right)^2$$

$$v(\omega, x_0, y_0) = \sqrt{K^2(\omega, x_0, y_0) - \omega^2 V^{-2}}$$

For the Fresnel wave the function $C(\omega, t_0)$ has the form

$$C(\omega, t_0) = \frac{\omega^2}{2 V K(\omega, x_0, y_0) v(\omega, x_0, y_0)} \left( \frac{\partial f(x_0, y_0, z_0, \omega)}{\partial z_0} \right)^2$$

We will write down the first term of the asymptotic form of the vertical velocity component of the global Airy wave resulting from the motion of a point mass source in a stratified horizontally inhomogeneous medium:

$$w = w_0(t, t_0, \omega) f(x, y, z, \omega) \, \text{Ai}' \left( \left[ \frac{3}{2} S^*(t, t_0, \omega) \right]^{2/3} \right) \tag{12}$$

$$w_0(t, t_0, \omega) = \frac{\omega^2 \, K(\omega, x, y) \, \sigma^{1/4}(\omega, x, y)}{[2 V K_\omega'(\omega, x, y) D(t, t_0, \omega) K^3(\omega, x_0, y_0) v(\omega, x_0, y_0)]^{1/2}} \frac{\partial f(x_0, y_0, z_0, \omega)}{\partial z_0}$$

where $x = x(t, t_0, \omega)$, $y = y(t, t_0, \omega)$. The first term of the asymptotic expansion for the rise of the global Fresnel wave has the form:

$$\eta = \eta_0(t, t_0, \omega) f(x, y, z, \omega) \, \Phi\left( \sqrt{2 S^*(t, t_0, \omega)} \right) \tag{13}$$

$$\eta_0(t, t_0, \omega) = \frac{\omega}{[2 V K_\omega'(\omega, x, y) \, K(\omega, x_0, y_0) v(\omega, x_0, y_0)]^{1/2}} \frac{\partial f(x_0, y_0, z_0, \omega)}{\partial z_0}$$

In a horizontally homogeneous medium the solutions (12)-(13) coincide with the global asymptotics constructed in [1-3]. For large $S^*$ (at points remote from the wave front), using the asymptotic forms of the Airy function and the Fresnel integrals for large values of the argument, we obtain the usual WKB expansion for a horizontally inhomogeneous medium, which coincides with that obtained in [4-10]; for small $S^*$ (at points close to the wave front) we obtain the solution in the weak-dispersion approximation [2]. Thus, in their most general form the solutions constructed describe the internal wave field associated with the motion of a source in a stratified horizontally inhomogeneous medium.